\documentclass[12pt]{article}
\pdfoutput=1
\usepackage{graphics, color}
\usepackage{graphicx}
\usepackage{amsmath}
\usepackage{amssymb}
\usepackage{amsfonts}
\usepackage[usenames,dvipsnames]{xcolor}
\usepackage[normalem]{ulem}

\definecolor{orange}{rgb}{1,0.5,0}
\definecolor{grey}{rgb}{.5,.5,.5}
\definecolor{bluegreen}{rgb}{0,.5,.5}
\definecolor{darkgreen}{rgb}{0,.5,0}

\newcommand{\sect}[1]{\section{#1}\setcounter{equation}{0}}

\def\gsim{\, \rlap{$>$}{\lower 1.1ex\hbox{$\sim$}}\,}
\def\lsim{\, \rlap{$<$}{\lower 1.1ex\hbox{$\sim$}}\,}
\newcommand{\be}{\begin{equation}}
\newcommand{\ee}{\end{equation}}

\begin{document}


\begin{titlepage}
\bigskip
\bigskip\bigskip\bigskip
\centerline{\Large \bf An Apologia for Firewalls}


\bigskip\bigskip\bigskip
\bigskip\bigskip\bigskip

 \centerline{{\bf Ahmed Almheiri,}\footnote{\tt ahmed@physics.ucsb.edu}*
 {\bf Donald Marolf,}\footnote{\tt marolf@physics.ucsb.edu}*
 {\bf Joseph Polchinski,}\footnote{\tt joep@kitp.ucsb.edu}*${}^\dagger$}
 \centerline{ {\bf Douglas Stanford,}\footnote{\tt douglasstanford@gmail.com}${}^{\dagger \ddagger}$\
 and {\bf James Sully}\footnote{\tt sully@physics.ucsb.edu}*}
\bigskip
\centerline{\em *Department of Physics}
\centerline{\em University of California}
\centerline{\em Santa Barbara, CA 93106 USA}
\bigskip
\centerline{\em ${}^\dagger$Kavli Institute for Theoretical Physics}
\centerline{\em University of California}
\centerline{\em Santa Barbara, CA 93106-4030 USA}
\bigskip
\centerline{\em ${}^\ddagger$Stanford Institute for Theoretical Physics }
\centerline{\em and Department of Physics, Stanford University}
\centerline{\em Stanford, CA 94305, USA}
\bigskip\bigskip\bigskip


\begin{abstract}

We address claimed alternatives to the black hole firewall.  We show
that embedding the interior Hilbert space of an old black hole into the Hilbert space of the early radiation is inconsistent, as is embedding the semi-classical interior of an AdS black hole into any dual CFT Hilbert space.  We develop the use of large AdS black holes as a system to sharpen the firewall argument.  We also reiterate arguments that unitary non-local theories can avoid firewalls only if the non-localities are suitably dramatic.

\end{abstract}
\end{titlepage}

\baselineskip = 16pt

\tableofcontents

\setcounter{footnote}{0}

\bigskip
\centerline{\sc Notation\footnote{\label{DP}This subject has suffered from conflicting notations, so we will try to standardize here.  Our notation for mode operators follows Ref.~\cite{Almheiri:2012rt}, but we now align the Hilbert space notation with that of the modes.  Thus, $E$, $B$, $\tilde B$ here were $A$, $B$, $C$ in \cite{Almheiri:2012rt} but are commonly called $C$ (or $R$), $B$, $A$ in much of the literature.  For example, the proposal $\tilde B \subset E$ investigated in \S2 is referred to elsewhere as $A \subset R$.
Note that in our notation $\pmb a$ is the natural set of modes that would be used by an infalling observer, commonly called Alice.}}

\indent

$b$, a late-time outgoing Hawking mode

$\tilde b$, the interior partner of $b$

$a$, a mode smoothly spanning the horizon

 $e_b$, an early-time Hawking mode, entangled with $b$

 $e$, a generic early Hawking mode

Here the terms early and late are relative to some chosen time after the Page time, where we take the latter to mean the time at which the von Neumann entropy of the emitted Hawking radiation first begins to decrease \cite{Page:1993df}.  To minimize notation, in each case above we use the given symbol to denote both the mode itself and the corresponding annihilation operator, with $N_a,N_b,N_i,N_{e_b},N_e$ the associated number operators.  Capital letters $B,\tilde B, A ,E_B, E$ will denote Hilbert space factors in which such operators act.  Boldface, e.g.\ $\pmb a, \pmb b$, denote collections of such modes. For example, the $\pmb a$ are linear combinations of $\pmb b, \pmb b^\dagger, \tilde {\pmb b}, \tilde {\pmb b}^\dagger$.

\section{Introduction}

In 1976, Hawking~\cite{Hawking:1976ra} observed that black hole radiation is produced in a highly mixed state, which leads at the end of the evaporation process  to either information loss or  a remnant.  This argument uses general relativity as an effective field theory only in regions of low curvature.  On the other hand, as we will review, AdS/CFT duality implies that Hawking radiation is essentially pure and therefore requires a premature breakdown of effective field theory.  The argument from duality is indirect, and does not explain how the bulk physics is modified.  In recent work~\cite{Almheiri:2012rt}, inspired by the unitary evaporation models  of Mathur
\cite{Mathur:2009hf} and Giddings~\cite{Giddings:2011ks,Giddings:2012bm}
and the quantum information perspective of Hayden and Preskill~\cite{Hayden:2007cs}, we have attempted to sharpen this issue.

Specifically, Ref.~\cite{Almheiri:2012rt} considered a late-time outgoing Hawking mode $b$, emitted after the Page time.  Under the seemingly mild assumption that infalling observers find no dramatic effects at the horizon, and in particular that they find only an exponentially small number of high energy particles, effective field theory requires that $b$ be highly entangled with a mode $a$ behind the horizon.  On the other hand, purity of the Hawking radiation requires that $b$ be entangled with the early radiation $E$.  This collection of mutual entanglements exceeds the limit allowed by quantum mechanics.\footnote{\label{GBfootnote}In particular, {taken together} {$\tilde{b}$ and $b$} are {in a pure state} {if we use} an eigenbasis of Schwarzschild frequency, and arbitrarily close to this for wavepackets that are narrow in frequency~\cite{Giddings:1992ff}, leaving no room for any entanglement with $E$.  Thus  the reduction of the {$b$-$E$ entanglement} by gray body factors {of order 1}, by entropy production during the radiation process~\cite{Page:1976ki,Zurek:1982zz,Verlinde:2012cy}, and by the possibility that the Page curve~\cite{Page:1993df} is not saturated do not help.  The basic entanglement conflict was well-discussed before our work; see e.g. \cite{Sorkin:1997ja,Mathur:2009hf,Braunstein:2009my,Giddings:2012dh}.  We note that  \cite{Mathur:2009hf,Giddings:2012dh} show explicitly that the conflict cannot be resolved by a slow build up of small effects.} It is assumed here that the propagation of the Hawking mode $b$ from the stretched horizon (set by some UV cutoff on the effective field theory) to the asymptotic region does not deviate from effective field theory by amounts of order one.  These three assumptions, No Drama, Purity, and EFT (outside the stretched horizon for macroscopic black holes), had been stated explicitly as postulates of Black Hole Complementarity (BHC)~\cite{Susskind:1993if}.  A fourth postulate states that the Bekenstein-Hawking entropy $S_{\rm BH}$ governs the black hole density of states for all purposes outside the stretched horizon, implying that the black hole is still macroscopic at the Page time as defined above.  We note that Ref.~\cite{Almheiri:2012rt} also made the explicit assumption that no observer will see any violation of quantum mechanics.  One of these assumptions, or some other that was unrecognized, must break down.

Guided by AdS/CFT duality, Ref.~\cite{Almheiri:2012rt} did not consider violations of purity, of quantum mechanics, or of the assumption that $S_{BH}$ governs the density of states.  It further argued that violations of EFT are strongly limited by consideration of mining experiments (see also \cite{Unruh:1982ic,Brown:2012un}) which allow the possibility of manipulating quanta as they leave the stretched horizon.
The remaining alternative, {that observers crossing the horizon experience drama in the form of a firewall} of high energy particles,\footnote{Discussions of the link between the UV structure of the quantum state and transhorizon correlations can be found in \cite{Haag:1984xa,'tHooft:1984re,Bombelli:1986rw,Sorkin:1986zj,Fredenhagen:1989kr,Srednicki:1993im,Jacobson:2003vx,Giddings:2006sj} and are particularly {quantitative and} explicit in \cite{Giddings:2011ks}.
See also Ref.~\cite{Braunstein:2009my}, which had previously argued that black hole evaporation would result in ``a loss of trans-event horizon entanglement" and thus ``fields far from the vacuum state in the vicinity of the event horizon,'' termed there an {\it energetic curtain.} However this argument was based on a model of black hole evaporation that explicitly violates EFT, allowing \cite{Braunstein:2009my} to conclude that the energetic curtain can be delayed until the black hole evaporates to the Planck scale.}  was offered as the simplest resolution.  In particular, mining experiments also imply that our issues extend beyond the low angular momentum modes that dominate normal Hawking emission and afflict all modes near the horizon. {Though it need apply only to sufficiently old black holes,\footnote{{The firewall must turn on before the late modes are emitted; i.e., at least around the Page time.  But Ref.~\cite{Almheiri:2012rt} noted reasons to believe it may become active much sooner, around the so-called fast scrambling time \cite{Hayden:2007cs,Sekino:2008he}.  The idea that old black holes may differ qualitatively from young black holes as defined by some timescale in this range was advocated  previously in \cite{Mathur:2008kg,Braunstein:2009my,Giddings:2011ks}.}} our conclusion is nevertheless radical,} and many authors have proposed alternatives.  We address many of these ideas below, and argue that {they are inadequate} in their current forms.

In \S2 we reconsider a proposal that has been made frequently: since $b$ is entangled with both $e_b$ and $\tilde b$, these must actually represent the same bit in a single Hilbert space: $\tilde B \subset E$.  While this is seemingly in keeping with the spirit of BHC~\cite{Susskind:1993if,Stephens:1993an}, we reiterate that it fails to respect one of the key constraints of BHC as formulated in~\cite{Preskill,Susskind:1993mu}.  That is, no observer should be able to see both copies of the bit, else their observations cannot be described by quantum mechanics.  We also sharpen this argument with a thought experiment where the cloned bit is sent to the infalling observer after he crosses the horizon.  Moreover, while the thought experiment in Ref.~\cite{Almheiri:2012rt} required an elaborate quantum computation, we show that a single crude measurement on the early radiation is enough to lead to observable violations of quantum mechanics and/or a firewall.  We also address the issue of the `state-dependence' of some proposed constructions.

In \S3 we discuss {the addition of
explicit nonlocal interactions, in particular the so-called nonviolent nonlocality scenarios of \cite{Giddings:2011ks,Giddings:2012gc,Giddings:2013kcj}.  We explain the difficulties inherent in using such effects to restore unitarity in black hole evaporation.}

Section 4 considers black holes in an AdS box.  We argue that this is a convenient way to see that center-of-mass drift~\cite{Page:1979tc}, suggested by Refs.~\cite{SussUn, Nomura2,Hsu} as a loophole in the argument of Ref.~\cite{Almheiri:2012rt}, is actually a red herring.
We  show that thought experiments in AdS evade limitations that might otherwise arise from the time scale of quantum computation~\cite{Harlow:2013tf}.
We also use AdS/CFT duality to exclude remnants, and to further sharpen the argument against nonviolent nonlocality.

Section 5 revisits the old problem of using AdS/CFT to write fields in the black hole interior in terms of CFT operators. Here we explore lines of reasoning independent of our original firewall argument.  In particular, we focus on black holes that reach equilibrium and do not evaporate.  We explain that the classic bulk propagation construction fails due to trans-Planckian effects for black holes older than the fast scrambling time.  Moreover, we argue  that such fields {\it cannot} be embedded in the Hilbert space of the \mbox{CFT}.  {Even outside of AdS/CFT, the same issue applies  to any attempt to describe the black hole interior in terms of a fixed Hilbert space of finite dimension $e^{S_{\rm BH}}$.}  We also give an argument, somewhat different in structure from that of Ref.~\cite{Almheiri:2012rt} but similar in its basic assumptions, that typical equilibrium black holes have firewalls.

In \S6 we consider another issue, the relation between the black hole state seen by the external observer, and the density matrix of the infalling observer.  This gives another perspective on issues discussed earlier, in particular emphasizing that the state-dependence of several constructions is a modification of quantum mechanics.  We argue that there is no satisfactory choice for the infalling density matrix that avoids the firewall.

We end with a discussion of further general issues in \S7.  We note that a key theme of our comments is the inconsistency of proposals allowing the release of energy by the black hole to be physically separated from the escape of information.

\sect{Problems with $\tilde B \subset E$}
\label{AE}

Since the basic problem is that the bit $b$ is required to be entangled with both $\tilde b$ and $E$, a natural way to try to save quantum mechanics is to suppose
that $\tilde b$ is actually contained in $E$.  That is, the interior Hilbert space $\tilde B$ of an old black hole is embedded in the larger Hilbert space $E$ of the early radiation.  This is in keeping with the original
interpretation of black hole complementarity~\cite{Susskind:1993if,Stephens:1993an,Preskill}, that the Hilbert space structure of quantum mechanics is unmodified, but locality breaks down radically so that different observers see the same bit in macroscopically different locations.
Recently this idea has been considered by a number of authors, including {Ref. \cite{Bousso:2012as} (which argued against it) and Refs.~\cite{Srednicki,Nomura:2012sw,Nomura2,Papadodimas:2012aq,Nomura3, Harlow:2013tf,Susskind:2013tg}.   We now point out several difficulties with this proposal.}

{\it 1. Violation of quantum mechanics.} As discussed in Ref.~\cite{Almheiri:2012rt}, this idea runs afoul of one of the basic consistency checks for complementarity: if a single observer can see both copies of a bit, then there is cloning, and quantum mechanics has broken down.  In the present context, Alice can remain outside during the early radiation and extract from $E$ via a quantum computation a bit $e_b$ that will be strongly entangled with the later Hawking bit $b$.  She then jumps into the black hole, capturing the entangled bits $b$ and $\tilde b$ as she goes, and so possesses in her laboratory three bits with no sensible quantum description.  Their entanglements violate strong subadditivity~\cite{Mathur:2009hf}.

Various authors, \cite{Nomura2,Nomura3,Harlow:2013tf,sch}  
have proposed that excitations exist at the horizon only if this quantum computation has been done, and not otherwise.  This conflicts with the fact that the entanglement of $b$ and $E$ exists whether or not the experiment is done, but one might imagine that there is some modified notion of complementarity that makes it possible.  In fact, we can exclude this by an operational argument.  That is, the infalling observer need not actually carry the bit $e_b$.  Emmy, who stays outside the black hole, can extract the bit $e_b$ and send its quantum state to Alice so that it overtakes her after she crosses the horizon; Emmy's measurement can be done at spacelike separation from Alice's horizon-crossing.  For this experiment, it is inconsistent with quantum mechanics for Alice to have seen vacuum at the horizon.

{\it 2. Perturbing $e$ creates a firewall.}  Ref.~\cite{Harlow:2013tf} has argued that the quantum computation cannot be done rapidly enough to allow this experiment.  Section \ref{inAdS} will address this issue directly, but here we provide a new and simpler argument  for the inconsistency of $\tilde B \subset E$.  Suppose that Bob attempts an experiment similar to Alice's, but not being in possession of a quantum computer he instead measures some simple early bit $e$, say a particular early Hawking photon.  We claim that the commutator of $e$ with $N_a$, the excitation above the infalling vacuum, is of order one.

For simplicity we work with the parity $(-1)^{N_e}$.  Let us take a basis in which
\be
(-1)^{N_e} = \sigma^z \otimes I \,;  \label{basis}
\ee
that is, we factor the Hilbert space into the measured parity and the rest.  Now consider $(-1)^{N_{{\tilde b}}}$.  Since  $\tilde B \subset E$, we may expand
\be
(-1)^{N_{\tilde b}} = I \otimes S^0 + \sigma^x \otimes S^x + \sigma^y \otimes S^y + \sigma^z \otimes S^z \,.
\ee
The matrices $S^\mu$ are constrained only by $(-1)^{N_{\tilde b}} (-1)^{N_{\tilde b}} = 1$.
The relation between the complementary descriptions is expected to involve a scrambling of the Hilbert spaces, so the operators $S^0$, $\vec S$ are generic and have eigenvalues of order one.  It follows that the commutator of $(-1)^{N_e}$ and $(-1)^{N_{\tilde b}}$ is of order one, and so therefore is $[e,\tilde b]$.

In particular, if we start with an eigenstate of $(-1)^{N_{\tilde b}}$ and measure $(-1)^{N_{e}}$, the eigenvalue of $(-1)^{N_{\tilde b}}$ changes with probability $O(1)$.  For convenience, we show this for the process with $\tilde b$ and $e$ switched, which is equivalent but clearer in the basis (\ref{basis}).  Let $(-1)^{N_e} |\psi\rangle = +  |\psi\rangle$.  Then
\be
\label{-1}
\langle \psi | (-1)^{N_{\tilde b}} (-1)^{N_e} (-1)^{N_{\tilde b}} |\psi\rangle
= \langle \psi |  \sigma^z \otimes( S^0 S^0 + S^z S^z - S^x S^x - S^y S^y )  |\psi\rangle + {\rm cross\ terms} \,.
\ee
We wish to average (\ref{-1}) over all $S^0,S^x,S^y,S^z$ consistent with $(-1)^{N_{\tilde b}} (-1)^{N_{\tilde b}} = 1$.
The cross terms involve products of distinct $S^\mu$ and so are on the average zero, since the constraint allows independent sign flips.  The distinct $S^\mu S^\mu$ are on average equal, so on average the expectation value is reduced from 1 to 0 by the measurement.

In terms of the infalling modes, $\tilde b$ can be expanded as a sum of $a$, $a^\dagger$, so the commutator of $e$ with one of these (generically both) is also of order one.
Now, suppose that there were no firewall, so that the infalling observer sees vacuum, $a |\psi \rangle = 0$.  After Bob measures his bit, the order one commutator $[e,a]$ means that the state has been perturbed.  This is true for every mode $a$, so Bob has created a firewall!  Of course, it is more natural to conclude that a firewall was there all along.  It may seem odd that measurement of a single bit can perturb many others, but this seems to be a manifestation of the butterfly effect: perturbation of a single bit, followed by a scrambling operation, perturbs all bits.

There is a possible subtlety here.  One measurement perturbs a second noncommuting measurement only if the latter is later in time.  For local field theories, there is an unambiguous time ordering because operators at spacelike separation commute.  Here we have to assign some foliation, and if $\tilde b$ is effectively `earlier' than $e$, the measurement of $e$ will not perturb it.  However, the infalling observer will encounter $e$ before $\tilde b$, so such a proposal would lead to a closed timelike loop.

{\it 3.  Not all electrons are physically identical.}  The commutator $[e,\tilde b]$ has another unpalatable effect.  Bob can capture $e$ without yet measuring it and, if there is no firewall, see effects of the nonzero commutator when he falls past the horizon.  On the other hand, he might carry instead a physically identical bit not captured from the early radiation (such as $b$), which commutes with $\tilde b$.\footnote{{Note that a simple observer follows a timelike path and so encounters these bits at timelike separation. Thus causality imposes no direct requirement that they commute.  In this context, the point is that all these
 bits are essentially outward-moving, functions of the Kruskal $U$, and so the commutator does not depend on the $V$ value at which we measure them: an $O(1)$ commutator $[e,\tilde b]$ is inconsistent with local field theory.  Of course, an observer can also send out probes to interact with spacelike-separated bits away from his worldline and then reassemble the results at a later time.}}
  So, letting the bits be electrons, we find that not all electrons are the same!  This issue also arises in other attempts to circumvent the firewall: linearity of quantum mechanics does not allow the {physics of} a bit to depend on either the bits history or on the degree to which it is entangled (though it can depend on the specific entanglement, e.g. two spin-$\frac12$'s can combine either to spin 0 or to spin 1).

{\it 4. State dependence.} The identification $\tilde B \subset E$ is based on the fact that these have the same entanglement with $B$.  However, the precise $E$-$B$ entanglement depends on the initial state of the black hole, and so this construction of the internal Hilbert space depends on this state in a problematic way {(see related comments in \cite{Bousso:2012as})}.  To see this, we follow Refs.~\cite{Papadodimas:2012aq,Verlinde:2012cy}, which have proposed explicit constructions of the Hilbert space of an infalling observer using the idea that it can be identified by its entanglement. These papers are mainly in the context of stable AdS black holes, but as those authors note the construction extends to evaporating black holes. Let $i$ index the space $I$ of initial black holes states, $j$ the states of the early radiation $E$ and $\pmb N$ the states of the late radiation $B$ in a Fock basis.  Given the black hole $S$-matrix $S_{i,j\pmb N}$, a particular initial state $i$ will decay as\footnote{{The discussion below focuses on the Hilbert space at future null infinity, and it directly applies only to the low angular momentum modes. A similar discussion can be applied to the higher $\ell$ modes, by working with the state at some point after the Page time but before complete evaporation.}}
\be
| i \rangle_I \to S|i\rangle_I = \sum_{j,\pmb N} S_{i,j\pmb N}| j, {\pmb N}\rangle_{E,B}  \,.
\ee
Defining
\be
|\tilde {\pmb N} \rangle_{\tilde B} \equiv {Z^{1/2}} e^{\beta E_{\pmb N}/2}  \sum_{j} S_{i,j\pmb N}| j \rangle_E \label{bine}\,,
\ee
the late-time state is
\be
{Z^{-1/2}}\sum_{\pmb N} e^{-\beta E_{\pmb N}/2} |\tilde {\pmb N}, {\pmb N} \rangle_{\tilde B, B} \,.
\ee
Identifying $|\tilde {\pmb N} \rangle_{\tilde B}$ as the Fock states of the interior Hawking modes, this is the infalling vacuum state as required by No Drama.\footnote{{In the language of Ref.~\cite{Papadodimas:2012aq}, the early and late Hilbert spaces are the fine and coarse Hilbert spaces.}}  Thus, the identification~(\ref{bine}) is the desired mapping from $\tilde B$ into $E$.

{Having identified the states $|\tilde {\pmb N} \rangle$, we can now define interior operators as linear combinations of the {$|\tilde {\pmb N}\rangle \langle
\tilde {\pmb N}' |$}.  For example, for the individual Hawking partner modes $\tilde b_k$ we have lowering and raising operators
{\be
\tilde b_k |\tilde {\pmb N} \rangle_{\tilde{B}} = \tilde{N_k}^{1/2} |\tilde{{\pmb N}} - \hat{{\pmb k}}  \rangle_{\tilde{B}}  \,,\quad
\tilde b^\dagger_k |\tilde {\pmb N} \rangle_{\tilde{B}} = (\tilde{N_k}+1)^{1/2} |\tilde{\pmb N}+  \hat{{\pmb k}} \rangle_{\tilde{B}} \,. \label{bbt}
\ee}
We have defined early and late so that the dimension of $E$ is much larger than $B$ and $\tilde{B}$.  As a result, states of the form (\ref{bine}) span a low-dimensional subspace of $E$ and  (\ref{bbt}) is an incomplete specification of $\tilde{b},\tilde{b}^\dagger$ as operators on $E$. One option, implicit in Ref.~\cite{Papadodimas:2012aq}, is to set all unconstrained matrix elements to zero. With this choice, we can fully define e.g. $\tilde b$ as
\begin{align}
\tilde{b}_k(i) &= \sum_{\tilde{\pmb{N}}}\tilde{N_k}^{1/2}|\tilde{{\pmb N}} - \hat{{\pmb{k}}}\rangle\langle \tilde{{\pmb{N}}}| \label{b3}\\
&=Z\sum_{\pmb{N}}e^{\beta (E_{\pmb{N}} + E_{\pmb{N} - \pmb{\hat{k}}})/2}N_k^{1/2}\langle \pmb{N} - \pmb{\hat{k}}|S|i\rangle\langle i|S^\dagger|\pmb{N}\rangle.\label{b2}
\end{align}
Here, ${}_B\langle\pmb{N} - \pmb{\hat{k}}|S|i\rangle_{I}$ is a ket vector in $E$, so for a given initial state the lower line manifestly maps $E\to E$.

We note that operators defined by the above prescription are
sparse in the space $E$. The calculation of $[\tilde b, e]$ in point 2 is thus slightly modified, but because $e$ is not sparse the conclusion again is that the effect is large.\footnote{We thank Daniel Harlow for discussions of this point.}

In addition, there is a new problematic feature in that the embedding of the interior Hilbert space in the early radiation depends on the initial state $i$: it is not just a mapping from $\tilde B \to E$, but from $ I \times \tilde B \to E$, where $I$ is the space of all initial states.
Consequently, operators in the interior become maps from $E \to E$ that depend on the reference state $i$.  This state-dependence is outside the normal framework of quantum mechanics, and one must argue very carefully to show that it is consistent; the points above suggest that it is not
(see also, e.g., \cite{Kapustin:2013yda}). We will return to issues of state-dependence in section \ref{intF}.

For clarity, we note that one should distinguish this from a more familiar form of state dependence, exemplified for example by the construction of bulk fields in AdS as a power series in bulk interactions~\cite{Kabat:2011rz,Heemskerk:2012mn}. A bulk field $\phi$ is constructed in terms of single-trace boundary operators ${\cal O}$ as
\be
\label{unexpanded}
\phi = K_1 {\cal O} + K_2 {\cal O}{\cal O}  + \ldots  \,,
\ee
where the $K_i$ are multilocal bulk-to-boundary smearing operators.\footnote{For fields localized in the angular directions, the smearing functions do not exist~\cite{Hamilton:2006fh,Papadodimas:2012aq,Bousso:2012mh}, but there is no problem working with partial waves.}
Expanding around a given background, so that ${\cal O} = \langle {\cal O} \rangle +\sigma$ and
$\phi = \langle \phi \rangle + \varphi$,
we have
\be
\label{vp}
\varphi = K_1 \sigma + 2K_2  \langle {\cal O} \rangle \sigma
+ K_2 \sigma\sigma + \ldots \,.
\ee
The particular expansion above depends on the background $ \langle {\cal O} \rangle$, so one might be tempted to call (\ref{vp}) `state-dependent'. However, the original definition (\ref{unexpanded}) makes it clear that all such expansions are consistent with a single expression as a linear operator. The state-dependence of (\ref{b2}) is not of this kind, as we will see in section \ref{intF}.

{\it 5. Arbitrariness.}  The construction of $|\tilde {\pmb N} \rangle_{\tilde B}$ is dependent on the choice of separation time between the early and late Hilbert spaces.  Any choice of more than half of the Hawking modes may be used to define an early/fine Hilbert space that has sufficient entanglement to embed $\tilde B \to E$ as above,  but each leads to a different embedding.

{\it 6. Energy considerations.} The operators $\tilde{b},\tilde{b}^\dagger$ defined in eq. (\ref{bbt}) change the energy of the early radiation, whereas the correct behind-the-horizon operators should change only the energy emitted at late times. Simple bulk observables, such as the gravitational field outside the horizon, will be sensitive to this distinction.

{\it Conclusion.} We conclude that $\tilde{B} \subset E$ is problematic.  One may ask if the situation is improved by weakening this strict inclusion.  For example, Refs. \cite{Nomura:2012sw,Nomura2,Nomura3} suggest that operators
$\tilde{b},\tilde{b}^\dagger$ may act non-trivially both on our $E$ and on a separate Hilbert space of black hole states.\footnote{In the language of \cite{Nomura:2012sw,Nomura2,Nomura3}, the operators $\tilde{b},\tilde{b}^\dagger$ depend on a branch of the wavefunction determined in part by the state in $E$.  This means that the full operators $\tilde{b},\tilde{b}^\dagger$ may be written as a sum of terms associated with different projection operators on $E$.  Thus the summed operator acts non-trivially on $E$.} To avoid objections 4 and 5, the operators should be state-independent. We might then parameterize the amount of action on $E$ by the expectation values of the commutators of some given $b, b^\dagger$ with the operators $e,e^\dagger$ associated with the early modes.  By point 2 above, at least one of these must be large if there is a significant commutator of
$b, b^\dagger$ with any qubit in $E$.  But then our objections 1, 3 and 6 apply directly. On the other hand, if $b, b^\dagger$ have small commutators with all qubits in $E$, then we may define slightly modified operators $c, c^\dagger$ that precisely commute with all qubits in $E$ and which define approximately the same notion of infalling vacuum as $b, b^\dagger$.  The result is again an entanglement conflict with unitarity.

Another potential alternative is the proposal~\cite{Bousso:2012as,Banks:2012nn,Harlow:2013tf,Susskind:2013tg} that there might be a stronger form of complementarity, in which there is no global Hilbert space.  Rather, each observer has their own Hilbert space, with suitable overlap conditions.  However, it remains to provide a working example that evades our arguments.  In particular, the most developed version~\cite{Banks:2012nn} requires the restrictions of the quantum states to agree in the observers' common causal past and thus appears to remain in direct conflict with \cite{Almheiri:2012rt}.

\sect{Problems with nonlocal interactions}
\label{noNL}

Even accepting our basic argument, one may hope to make violations of EFT benign.  This is, in particular, the goal of so-called nonviolent nonlocality (NVNL) models \cite{Giddings:2011ks,Giddings:2012gc,Giddings:2013kcj} that {at least for sufficiently old black holes} modify effective field theory by adding nonlocal interactions in the black hole exterior, {extending to $r - r_{\rm S} = O(r_{\rm S})$ (or, more generally, to some other power of $r_{\rm S})$. This was the region called `the zone' in \cite{Bousso:2012as,Susskind:2012rm}, and one might say that the idea of NVNL is to make the information transfer nonviolent by completing the transfer only as the relevant mode passes through the zone. Thus the modes remain} in the infalling vacuum until, as measured in an infalling frame, their frequency falls to of order the black hole temperature $T$.   But our original work \cite{Almheiri:2012rt} described problems with such scenarios.  We now make these very explicit by studying a particular example.\footnote{{Although this model is directed at proposals for nonlocal interactions, the idea that information  `jumps over the zone',  may also characterize other attempts to evade the firewall.  We thank Daniel Harlow for discussions of this point.}} Here we focus on issues related to the evaporation itself, though since the model posits a globally-defined Hilbert space it will also suffer from the counting problems to be described in section \ref{intF}, which forbid such a Hilbert space.

Consider a product basis {adapted to} an outgoing bit $b$, its Hawking partner $\tilde b$, and {a further separation of} the rest of the black hole Hilbert space into a particular bit $h$ and the rest $H'$:
\be
|i\rangle_{H'} |j\rangle_{h} |k\rangle_{\tilde b} |l\rangle_{b} \,.
\ee
When $b$ first leaves the stretched horizon, it is entangled with $a$:\footnote{We neglect the Boltzman factors in what has become known as the qubit-model approximation. \label{qubit}}
\be
\sum_k  |i\rangle_{H'} |j\rangle_{h} |k\rangle_{\tilde b} |k\rangle_{b} \,. \label{hawkstate}
\ee
Now imagine that as $b$ moves though the {zone}, rather than propagating freely, there is an exchange between $b$ and $h$, so that when $b$ leaves the {zone}, a generic state has evolved according to
\be
\label{ti}
|i\rangle_{H'} |j\rangle_{h} |k\rangle_{\tilde b} |l\rangle_{b}
\to
|i\rangle_{H'} |l\rangle_{h} |k\rangle_{\tilde b} |j\rangle_{b} \,.
\ee
The Hawking state~(\ref{hawkstate}) thus evolves to
\be
\sum_k  |i\rangle_{H'} |k\rangle_{h} |k\rangle_{\tilde b} |j\rangle_{b} \,. \label{leavezone}
\ee

This has the desired effect that the entanglement with $\tilde b$ is transferred from $b$ to $h$, while the entanglement with the early radiation is transferred from $h$ to $b$.  If the black hole starts with $N$ bits in a generic state (that is, $N-1$ bits in $H'$ and one in $h$), in the Hawking state~(\ref{hawkstate}) it has $N+1$ bits of information, while in the final state~(\ref{leavezone}) the internal entanglement means that black hole retains only the $N-1$ bits of information in $H'$.

This is certainly an odd model: the Hawking process that produces the entangled bits has nothing to do with the bit $|j\rangle_{b}$ that eventually emerges from the {zone}.  Rather, the Hawking bit has been reabsorbed by the black hole, while the information characterized by $j$ has effectively jumped over the {zone}.

However, even this oddity does not solve the problem.  Via the mining procedure of \cite{Unruh:1982ic,Brown:2012un}, we can interact with the bit $b$ while it travels through the {zone}, for example introducing a phase:
\be
\sum_k  (-1)^k |i\rangle_{H'} |j\rangle_{h} |k\rangle_{\tilde b} |k\rangle_{b} \,.
\ee
{The introduction of this phase requires no exchange of energy, and so can be performed without entangling our system with the mining equipment.\footnote{This improves on the argument of \cite{Almheiri:2012rt}, which also considered processes that exchange energy.}}
The final state is now
\be
\sum_k (-1)^k   |i\rangle_{H'} |k\rangle_{h} |k\rangle_{\tilde b} |j\rangle_{b} \,.
\ee
In this way we can reach a larger set of black hole states, $N$ bits worth rather than $N-1$.  At the same time, the emission of radiation has lowered the mass of the black hole, so we have a process that allows the black hole to evaporate without releasing information.  The number of its internal states will then exceed that allowed by the Bekenstein-Hawking entropy\footnote{One can avoid a contradiction with unitarity by supposing that instructions given to the mining equipment to act on $b$ also affects the dynamics encoded in  (\ref{ti}).  However, required evolution is one that first undoes the unitary applied to $b$ above before swapping $h$ and $b$.   Although linear, this seems highly implausible.  We thank Yinbo Shi for discussions on this point. \label{implausible}}.

This is a prototype for one class of models in which the Hawking process is modified by nonlocal interactions, so-called nonviolent nonlocality.  More generally~\cite{Giddings:2011ks,Giddings:2012gc,Giddings:2013kcj} one might retain the original Hawking emission, which emits energy while increasing the entanglement of the black hole with the radiation, and introduce a second process which reduces the entanglement rapidly enough to offset this.

However, this second process is difficult to implement. It requires emitting some energy $\Delta E$.  The black hole has some temperature $T$, and so by the first law this lowers its Bekenstein-Hawking entropy by $\Delta E/T$.  To avoid making the entanglement conflict even worse than before, the black hole's entanglement must decrease even more, by some $\Delta S > \Delta E/T$. Thus the new process reduces the free energy $F = E-TS$ of the modes to which it couples.  Considering for the moment a simplified black hole for which the grey-body factor of any mode is either one (perfect transmission) or zero (perfect reflection), we see that it can make no use of the transmitted modes.  The original Hawking effect populates such modes with occupation numbers given by the Gibbs ensemble, with relative weights $e^{-N\omega/T}$, which already minimizes their free energy.

It follows that any successful use of the original Hawking modes for this purpose relies on the grey-body factors.  This is essentially equivalent to saying that it involves those parts of the modes that impinge on the black hole from infinity and reflect off of the angular momentum barrier.  Let us therefore continue to model the grey-body factors as zero or one and instead attempt to make use of those modes that perfectly reflect.
Thus, we imagine a process which transports quanta from behind the horizon past the potential barrier where they become outgoing quanta of large $L$.  A problem, however,  is that these same modes may also be populated by reflection of the incoming high-$L$ quanta.  While there are several ways one might try to deal with this, each remains unsatisfactory. For each option below, we consider a black hole with density of states $e^{S_{BH}}$ already maximally entangled with some reference system.

\begin{enumerate}
\item
Overwrite the incoming state.  Clearly, this is not unitary.  This is essentially the same issue faced in our model (\ref{ti}) above.
\item
Transport the incoming quanta into the black hole.  The problem is that this allows the reverse process, to pump information into the black hole at precisely the same rate at which it is emitted in the high-$L$ mode.  When combined with the effect of the original Hawking emission, the black hole's entanglement would then grow to exceed its density of states.
\item
Combine two channels into one: upshift the incoming quanta by some fixed frequency $\omega_0$, and use the remaining low frequency modes to carry out the information.  The problem is that each incoming quantum takes energy $\omega_0$ from the black hole without reducing entanglement, and so by stimulating this process we can reduce the Bekenstein-Hawking entropy below the entanglement entropy.
\item
Combine two channels into  one using the IR rather than the UV: clean the incoming mode by transporting the incoming quanta to a new outgoing mode a distance $R$ or more further away from the black hole, and use the mode cleared out in this way to carry the information.  The problem is that, in order to avoid a problem like 1 above, we must also clean this new outgoing mode by transporting its incoming quanta even farther away, etc.  Thus the non-locality acts arbitrarily far from the black hole.  Another way of seeing the same result is to note that this mechanism cannot succeed in the presence of an IR cutoff.
\item
Leave the incoming quanta in the given mode and simply take the new quanta to increase the occupation numbers.  This is of course impossible for fermions.  For bosons, the problem is similar to 3 above.  Adding a qubit to a given mode requires increasing by one the number of bits characterizing the occupation number of the mode; i.e., it requires doubling the energy of the typical incoming signal in this mode, draining significant energy from the black hole.
\item
Use modes that are purely outgoing.  One would need black holes that have a net excess of outgoing modes to avoid the issues above.  Generic black holes do not have such an excess, and likely there are no physical examples.\footnote{An unphysical example is a magnetically charged black hole, for which charged chiral fermions have $|q|$ one-way $L=0$ modes whose direction depends on the sign of the charge $q$,  in a gauge theory with a Tr $U(1)$ anomaly.}
\end{enumerate}
More generally, for any given process one might design a mirror to reflect back only the extra quanta, leaving only the Hawking quanta and the original paradox.  Section \ref{RNL} will discuss using the boundary of AdS as such a mirror, emphasizing that it admits a UV-complete description via AdS/CFT. Summarizing our discussion above, a general lesson appears to be that scenarios which physically separate any transfer of energy from the transfer of information are inconsistent with a Bekenstein-Hawking density of states.

For completeness, we also mention another problem with (or feature of) such extra flux models.  If Hawking evaporation is a thermal process at the Hawking temperature $T$, the rate at which energy is emitted is governed by the total emission cross section, which necessarily agrees with the total absorption cross-section.  Changing the absorption cross section by a factor of order unity for modes with frequency of order $T \sim 1/M$ must significantly change both the quasi-normal mode spectrum of the black hole and the gravitational waveform produced when two compact objects collide to form a black hole. {If astrophysical black holes are sufficiently old for this feature to be active,  the} resulting discrepancies with Einstein gravity should be visible to gravity wave interferometers in the near future.  Even allowing the emission to be non-thermal, similar comments apply to scenarios 2,3,4, and 5 above independent of the issues already discussed.  They may also apply to other proposals, e.g.\ \cite{Avery:2012tf}, that change the dynamics of such gravitational waves outside the horizon\footnote{We thank Borun Chowdhury and Steve Giddings for discussions of this point.}.

\sect{Evaporating AdS black holes}

\label{inAdS}

As noted in \cite{Almheiri:2012rt}, the cleanest version of our argument takes place in anti-de Sitter space, just as for the original information paradox~\cite{Maldacena:2001kr}.
There are at least two advantages to this formulation.  First, the thermodynamic stability of large AdS black holes suppresses fluctuations (one might say that it ``stabilizes moduli") and allows one to remain close to a single semi-classical background throughout the evaporation.  This eliminates certain criticisms of e.g.\ Refs.~\cite{Nomura2,Hsu}.  Second, as we review below, one may allow both energy and information to be transferred from the AdS spacetime to some intrinsically non-gravitational auxiliary quantum system which we introduce by hand.  The auxiliary system is then essentially unconstrained and so suffers no limitations on its ability to rapidly process quantum information.  Thus the concerns of Ref.~\cite{Harlow:2013tf} do not apply. While at the semi-classical level the above goals can also be achieved by placing an asymptotically flat black hole inside a perfectly reflecting box, the AdS context has the advantage that it admits a microscopic (i.e., UV-complete) description of such processes via AdS/CFT.  Below, we also take advantage of this opportunity to give a clean argument that the Hawking radiation is indeed pure in AdS/CFT so that, at least in that context, the scenarios advocated in \cite{Hossenfelder:2012mr,Jacobson:2012gh} cannot hold.

\subsection{Boundary conditions and couplings}
\label{AdSAMPS}

We wish to consider Schwarzschild-AdS black holes with radius $r_0 > \ell$ where $\ell$ is the usual AdS length scale.  Such black holes lie above the Hawking-Page transition and minimize the free energy at their temperature.  They also maximize the (generalized) entropy at fixed energy.  As a result, with the usual energy-conserving boundary conditions such black holes do not evaporate.  Instead, they quickly come into thermal equilibrium with a bath of their own Hawking radiation.  Since the coupling in the bulk will be weak,  each emitted Hawking quantum reflects off the boundary and is then reabsorbed by the black hole.

But we may also consider AdS boundary conditions that allow energy to leave the AdS space.  Much useful technology for doing so has been developed by the AdS/CFT community \cite{Witten:2001ua,Berkooz:2002ug,Gubser:2002vv}, though it refers only to semiclassical AdS physics.  While applications of this technology are easily interpreted in terms of any dual CFT, the technology can be used without assuming such a duality to hold.

Let us consider a theory of asymptotically AdS gravity with a tachyonic scalar field $\phi$ whose negative $m^2$ lies in the so-called Breitenlohner-Freedman window
$-d^2/4 < m^2 < -d^2/4 +1$.  Near the AdS boundary solutions of the linearized field equations admit the asymptotic expansion
\begin{equation}
\phi (x,z) = \alpha(x) z^{\Delta_-} +  \beta(x) z^{\Delta_+} + \dots,
\end{equation}
where $x$ represents a set of coordinates on the AdS boundary, $z$ is a so-called Fefferman-Graham radial coordinate, $\Delta_\pm = d/2 \pm \sqrt{d^2/4 + m^2}$, and the dots represent higher order terms in $z$. While either $\alpha(x)$ or $\beta(x)$ (or both) can be used to specify boundary conditions, we shall focus on $\beta(x)$ below.  Perhaps the simplest energy-conserving boundary condition is then to set $\beta(x) =0$, while fixing $\beta(x)$ to be some time-dependent function gives a boundary condition that can add energy to (or remove energy from) our AdS system.  In terms of any dual CFT, choosing a fixed but non-zero $\beta(x)$ is equivalent to coupling the CFT to a similar source for the CFT operator ${\cal O}$ dual to the bulk field $\phi$ in the $\beta=0$ theory.  We will therefore refer to such AdS boundary conditions as coupling the bulk scalar field to a boundary source.  The bulk variational principle compatible with such boundary conditions contains a term of the form
\be
S_{\rm bulk \, source} = \int_{\rm bndy} \sqrt{\gamma} \,\alpha(x) \beta(x) d^dx\,,
\ee
where $\sqrt{\gamma}\, d^dx$ is an appropriate volume element on the AdS conformal boundary.  This term is dual to the coupling \be
S_{\rm CFT \, source} = \int \sqrt{\gamma}\, \beta {\cal O}
\ee
 as just described, in any dual CFT.

Below, we will choose somewhat more complicated boundary conditions that may be interpreted as coupling the AdS system to additional dynamical degrees of freedom.  This can be achieved by promoting the boundary source {$\beta$} 
to a dynamical variable while also adding by hand some independent dynamics for this new field; {see e.g.\cite{Witten:2003ya,Rocha:2008fe,Compere:2008us,Domenech:2010nf,Faulkner:2010tq,Andrade:2011dg} for related constructions}.

As a simple example, neglecting $S_{\rm bulk \,source}$ we could take $\beta$ to be a free massive scalar on the AdS boundary.  In more complicated scenarios $\beta$ could be some composite field built from a large number $M$ of fields on the AdS boundary, {or even a field propagating in a higher number of dimensions~\cite{Rocha:2008fe}.  In either case the effect on any CFT dual is just to couple the CFT to this new sector using $S_{\rm CFT \, source}$.  The coupled theory is well-defined in the UV when the coupling is a relevant perturbation of the decoupled theory.\footnote{While the dimensions of $\beta$  and ${\cal O}$ must therefore sum to less than $d$,   this is easily achieved e.g. for AdS${}_4$ ($d=3$) by taking $\phi$ to have $m^2 = -2/\ell^2$
 so that, interpreting our model as a truncation of 11-dimensional supergravity on AdS${}_4$ $\times S^7$, the dual CFT operator ${\cal O}$ is the dimension 1 operator of the ABJM theory \cite{Aharony:2008ug}.}  If both the original CFT and the additional auxiliary system are stable, one expects that the Hamiltonian of the coupled system remains bounded below when the coupling is sufficiently weak\footnote{A general proof of this statement is not available using bulk methods, but such reasoning is supported by bulk results \cite{positive} which indicate that small sufficiently relevant deformations leave the Hamiltonian bounded below.}.

We will be interested in time-dependent such couplings.  In the above language, this may be achieved by making the $\beta$-dynamics time-dependent in such a way that passing to some canonically-normalized version {$\beta_{\rm can}$} renders the action independent of time except for the single term $S_{\rm bulk \, source}$ which now becomes
\be
g(t) \int_{\rm bndy} \sqrt{\gamma} \beta_{\rm can}(x) \alpha(x) d^dx
\ee
 in terms of some time-coordinate $t$ on the AdS conformal boundary. Since $g(t)$ encodes the only remaining time-dependence, the total energy is conserved when $g$ is constant and changes only very slowly if $g(t)$ varies adiabatically.  When $g(t) =0$, the system reduces to the original asyptotically AdS gravitational system together with an auxiliary decoupled {$\beta_{\rm can}$-system}, which we will henceforth refer to as AUX. A related scenario was recently discussed in \cite{Avery:2013exa}.

\subsection{Purity and Drama in AdS}

With this technology and notation in hand we can now state the scenario of interest.  Suppose that $g(t)$ vanishes in the far past and that the system begins in its ground state; i.e., the bulk spacetime is precisely empty AdS space and AUX is in its own non-degenerate ground state.  At some time, we turn on a non-dynamical time-dependent boundary source for our bulk scalar $\phi$, pumping scalar radiation into the bulk.  For sources of large enough amplitude this radiation will collapse gravitationally to yield a large AdS black hole.  At this point, we turn off the boundary source and the black hole settles down to a very large Schwarzschild-AdS black hole (with $r_0 \gg \ell$) in equilibrium with its Hawking radiation.

 We then take $g(t)$ to become non-zero.  Let us suppose that $g$ saturates at some small value $g_0$ for which the coupling to AUX may be considered weak.  We also take the density of states in AUX to be very large compared with that of the AdS system (e.g., $M \gg N^{3/2}$ if $N$ is the rank of the gauge group in an ABJM theory dual to an AdS${}_4$ bulk).  Then at least in the semi-classical approximation our AdS system behaves like a hot rock radiating into empty space, whose role is played here by AUX~\cite{Rocha:2008fe}.  As energy flows from the AdS spacetime to AUX, our AdS spacetime cools.  But since $g_0$ is small, the cooling is slow.  The AdS system has time to thermalize and remains well-described by a large, though now {somewhat} smaller, black hole in equilibrium with a bath of Hawking radiation.   We then return the coupling $g(t)$ to zero after some long time that has allowed most of the energy to escape, but which leaves us with a final black hole of size $r_0 > \ell$ so that we remain always in the stable regime.

{Since large AdS black holes are thermodynamically stable, fluctuations} of the center-of-mass position, total mass, and all other properties of the black hole remain small.  It has been suggested~\cite{SussP,Nomura2,Hsu}
that center-of-mass drift might complicate the necessary measurements of the black hole, but for a large AdS black hole one can readily verify that both the quantum and thermal spread of the center-of-mass distribution are {suppressed by a power of} the Planck length. In particular, a harmonic oscillator model gives center of mass position fluctuations of {order
\be
\Delta x \sim \ell {\left( \frac{\ell_p}{T\ell^2} \right)^{\frac{d-1}{2}}}
\ee
which is $\le \sqrt{\ell \ell_p}$ for $d \ge 2$ }where $\ell_p$ is the $(d+1)$-dimensional Planck length.  We should also point out that even without stabilization in AdS, drifts in the center-of-mass, total mass, and other properties can  be accounted for simply by measuring the corresponding properties of the outgoing photons.  It is worth noting that the entropy in the distribution of such macroscopic quantities is at most logarithmic in the mass, negligible compared to the overall entropy.

It is now straightforward to apply the argument of \cite{Almheiri:2012rt}.  We make four assumptions: i) The full (coupled) system is described by standard quantum mechanics with a (globally defined) space of states.
ii) The initial state of the joint system described above (pure AdS bulk, AUX vacuum) is indeed pure. iii) the full (coupled) system evolves unitarily. It thus remains pure.  iv) So long as the AdS black hole remains large, the entanglement of the AdS system with AUX is always bounded by the Bekenstein-Hawking entropy up to small corrections.  These assumptions are directly implied by the standard formulation of AdS/CFT,\footnote{Here we take as given that the CFT density of states is finite and agrees with the bulk Bekenstein-Hawking entropy above the Hawking-Page phase transition.  This is indicated by the striking agreement~\cite{Hanada:2008ez} of numerical simulations in analogous but slightly more complicated models. } and also more generally as we discuss below.  Of course, we also assume that standard low-energy effective field theory describes both experiments performed far outside the black hole and (with No Drama) by any infaller.

To proceed, consider a Hawking mode $b$ emitted from the black hole which, at some time, is in the process of being absorbed by AUX.  For the reasons explained in \S2, we take $b$, the interior partner mode $\tilde b$, and AUX to be independent systems.  This rationale will be further strengthened below.

For an infaller to find no drama, $b$ must be highly entangled with $\tilde b$.   But this prohibits significant entanglement with AUX.  Since the $b$-system is itself highly mixed, absorption by AUX necessarily increases the associated von Neumann entropy $S_{\rm AUX}$ by an amount of order $1$.  Here the weak coupling $g_0$ is analogous to increasing the effect of grey body factors which, as in footnote \ref{GBfootnote}, have no effect unless they are parametrically large.  So we take $g_0$ weak, but not parametrically so.  Note that AUX is directly coupled only to the particular bulk field $\phi$, and that the coupling of $\phi$ to other bulk fields is parametrically weak.  Thus all significant changes in the entanglement of AUX can be ascribed to its absorption and very occasional emission of $\phi$-quanta.\footnote{Turning the coupling between the CFT and AUX on and off generates some additional entanglement.  As long as the time scale for this at least the AdS light-crossing time, this additional entanglement entropy is $O(1)$ and does not affect the argument.}

It follows that at each time $t$, 
{the entanglement entropy between AUX and the CFT}
 is of the same order as $S_{\rm BH}({\rm initial}) - S_{\rm BH}(t)$.  Here $S_{\rm BH}({\rm initial})$ is the Bekenstein-Hawking entropy of the AdS black hole just before $g(t)$ becomes non-zero.   But once the black hole has lost most of its entropy this contradicts assumption (iv), that the entanglement is bounded by $S_{\rm BH}(t)$.  Unless prevented by some breakdown of effective field theory at low energies, any infaller will find drama at the would-be horizon.

In this form, our argument rules out the scenario presented in \cite{Jacobson:2012gh} which claimed that the contradiction in \cite{Almheiri:2012rt} can be avoided by taking the Hawking radiation itself to remain highly mixed, even at late time, and to be purified only by non-radiative degrees of freedom. But since AUX interacts only with the single bulk field $\phi$, it can be purified only by the absorption of $\phi$-quanta.  Assumptions (iii) and (iv) then imply that the Hawking radiation at late times is indeed strongly entangled with the radiation emitted at early times; i.e., purity holds in AdS/CFT.   Any more nebulous bulk degrees of freedom have no effect.  The proposal of \cite{Hossenfelder:2012mr}, involving an unobservable degeneracy in the Unruh vacuum, and the relevance of additional observables of the sort described in \cite{Gambini:2013ooa} are also ruled out (though they would in any case  raise concerns about the thermodynamics of the emitted radiation). Again we see that scenarios that physically separate the transfer of information from the transfer of energy are inconsistent with the black hole having a finite density of states.

Note that while assumptions (i-iv) follow from the standard formulation of AdS/CFT, they are in fact much weaker.  For example, we make no assumption that the algebra of observables on the AdS boundary is complete.  We therefore allow arbitrary superselection sectors in the sense of \cite{Marolf:2008tx,Marolf:2012xe}.  Since the superselected labels carry no energy, starting in the ground state of the AdS sytem forces the CFT to be in its (unique) ground state as well.  {There is no initial entanglement with the superselected labels, and none can be generated by the evolution since the labels do not couple to other degrees of freedom.} Thus the superselection sectors have no effect.

\label{comp}
\subsection{Return to $\tilde B \subset E$}

{Let us now strengthen the argument that $b$, $\tilde b$, and AUX are all independent systems. First note that, in the context of the AdS construction, the conjecture $\tilde B \subset E$ would imply that the interior Hilbert space of the AdS black hole actually lies in degrees of freedom entirely outside the AdS space -- a situation described in~\cite{Avery:2013exa} as `rather bizarre.'  There is in fact a sharp contradiction here.  If $\tilde B \subset \it AUX$, then any associated operators $\tilde b, \tilde b^\dagger$ commute with all CFT operators.  Both the total energy inside the AdS system (i.e., the energy captured by the ADM-like boundary term) and the total stress-energy in linearized fields outside the black hole are clearly dual to such CFT operators.  The former is just the CFT Hamiltonian, and the latter can be written in terms of CFT operators using constructions along the lines of \cite{Kabat:2011rz,Heemskerk:2012mn}.  Thus the difference, the total energy inside the black hole, is also a CFT operator that must commute with $\tilde b, \tilde b^\dagger$.  But this is impossible:  since adding a Hawking particle raises the energy outside, energy conservation requires adding a Hawking partner particle to lower the energy inside.}

{As a second point, we now} address the conjecture of \cite{Harlow:2013tf} that issues of computational complexity prohibit in principle the extraction from the early radiation of the particular qubit entangled with a late-time outgoing Hawking mode $b$, or at least prevent this qubit from being handed over to an infalling observer.  In our AdS scenario, this amounts to performing a quantum computation on AUX, which may of course be accomplished by a quantum computer also outside AdS (and perhaps part of AUX itself). By outsourcing the computation in this way we avoid any possible limits from complexity.  In particular, at a time of our choice we may set $g(t)=0$, decoupling AUX from the AdS system.  We may then accelerate the dynamics of AUX (and any quantum computer to which it is coupled) by simply multiplying their Hamiltonian by some acceleration parameter $\Upsilon (t) \gg 1$.  By taking $\Upsilon$ sufficiently large, we can perform any desired quantum computation in an arbitrarily short coordinate time $t$.  We then return $\Upsilon (t)$ to $1$ and turn on a coupling to the AdS space that hands the specified qubit to the desired infaller.  Equivalently, we may slow down the evolution in the CFT, $H_{\rm CFT} \to H_{\rm CFT} / \Upsilon (t)$.  Ref.~\cite{Harlow:2013tf} objects to this procedure, but it is in keeping with the standard rules of AdS/CFT duality~\cite{GKP,W}, which allow arbitrary deformations of the CFT Hamiltonian.  It is generally accepted that a black hole in AdS, with arbitrary boundary conditions, has the same information loss properties as a black hole in Minkowski spacetime; why should the same not apply to the experience of the infalling observer?

Much the same effect can be achieved keeping $\Upsilon(t) =1$ using the fact that for $g(t)=0$ AdS is a perfectly reflecting box.  We simply study a mode $b$ far enough in the future for our quantum computer to have extracted {\it any} desired qubit from AUX.  The fact that $b$ lies (perhaps exponentially) far in the future is no obstacle; we simply assume that the time evolution of the theory has been solved and computed some time in the distant past before the experiment takes place, and that the outcome has been used to design the particular quantum computer that will interact with AUX.   {Here we have in mind that we have some way protect our infaller from any perhaps-violent processes that the black hole may undergo on such long timescales, such as transporting him outside the AdS space before turning off the coupling to AUX.} This scenario is also possible in flat space if one allows the existence of perfectly reflecting boxes in which to place the black hole.

While we have argued that physical limits on quantum computation are not relevant for a black hole in an AdS box, we comment here on the limitations studied in Ref.~\cite{Harlow:2013tf} for a physical realization of an IR  AdS geometry on branes.  It was argued that, while the redshift between the IR black hole and the asymptotic region makes rapid computation possible, it is not possible to efficiently send the result  from the asymptotic region  to the IR.  We note here a different strategy that allows this.

Consider dangling a string from the asymptotic region into the IR. {Its energy is minimized when it is radial, so if we} move it sufficiently slowly, it should remain radial.  We can therefore use it like a pen: we write a message with its upper end, and the lower end will trace it out.  {In fact, this is consistent with the appendix to~\cite{Harlow:2013tf}. Applying} outgoing boundary conditions to the lower end, the resulting signal $\theta(\rho)$ is essentially independent of their radial coordinate $\rho$ for $\omega^2 R^2 \ll \rho \ll 1$, ranging from the IR well into the asymptotic region, indicating that the pen behaves as desired.  The difference is that Ref.~\cite{Harlow:2013tf} considers messages sent in from even farther away ($\rho \sim 1$), while we imagine holding the pen at smaller values of $\rho.$  Likely this strategy would work for other communication modes as well.

We note another possible limitation on communication.  If we accidentally move the pen too quickly or otherwise fail to isolate the black hole from the computation, we will send a signal that dwarfs the IR energy scale and destroys the coherence. Again, this is a concern for all modes that we might use to communicate.  It is necessary to isolate the upper end of the pen from the higher frequencies used to carry out the computation.
This does not seem to be a fundamental limitation:
leakage is generally exponentially small in the thickness of shielding, so a thickness of order $\ln \omega R$ should provide the desired isolation.

\subsection{Return to NVNL}
\label{RNL}

Our AdS thought experiment allowed AUX to receive information only from quanta of the AdS bulk field $\phi$.  In order to further sharpen our concerns regarding
nonviolent nonlocality \cite{Giddings:2011ks,Giddings:2012gc,Giddings:2013kcj}, we may further restrict the channel through which such information can flow. As a simple example, we may replace the bosonic field $\phi$ above with some fermion $\psi$.  If AUX couples to the AdS system only through $\psi$, then the Pauli exclusion principle guarantees us a finite channel-capacity per mode of $\psi$. Inserting a third system (the filter) between AUX and the AdS system and taking it to transmit efficiently only those modes of $\psi$ with frequency $\omega \sim T$ then forces the information to flow through a small set of modes already well-populated by the Hawking effect.  The only available NVNL scenarios are then very similar to the model described by (\ref{ti}) and are susceptible to the same mining argument.

\subsection{Remnants}

Remnants (see e.g.~\cite{Ori:2012jx} for a recent discussion) and other ideas such as \cite{Ashtekar:2005cj} that lead to information being recovered only long after the Page time are similarly incompatible with AdS/CFT; these would otherwise be an alternative to the firewall.\footnote{Ref.~\cite{Kim:2013fv} considers complementarity in the context of 2d gravity.  This theory leads to remnants~\cite{AS}, and so evades the firewall.}  We define a remnant as an object for which Hawking's original calculation is valid in regions of small curvature, but where the dynamics may change when curvatures are of order the Planck length.

Consider a large black hole weakly coupled to an exterior CFT as above.  Let us throw entangled bits in from the AdS boundary at a rate that just offsets the radiation of energy outward.  Since the curvature is small, by assumption the entanglement of the black hole with the exterior theory grows, and without bound.  This would require the black hole CFT to have an unbounded number of states below some finite energy.  The understanding of gauge theories is sufficient to exclude this.\footnote{This assertion is clearer for superrenormalizable duals such as the D2-brane, for which a lattice construction is more explicit.  Note that it is sufficient to consider the system at large but finite $N$: we are interested in black holes much larger than the Planck length, but the ratio does not need to be infinite.}

Ref.~\cite{Strominger:2009aj} has proposed `nonlocal remnants' as an alternative.  These decay more rapidly than the usual remnants because they are assumed to radiate not just into the $s$-wave but into all partial waves.  Nonlocal physics is required for a small object to decay into high partial waves.  Thus these are outside the spirit of remnants as usually considered, where deviations from effective field theory happen only in regions of large curvature.  Instead, they correspond in the terminology of Ref.~\cite{Almheiri:2012rt} to a violation of Postulate 2, effective field theory outside the stretched horizon.  In the specific scenario of Ref.~\cite{Strominger:2009aj}, the nonlocal physics is assumed to begin to operate only when the black hole reaches the Planck size, and so it would still be excluded by the boundedness argument above.

\sect{Static AdS black holes}
\label{intF}

While evaporating black holes raise very sharp problems for reconciling
Purity, EFT, and No Drama, it is enlightening to consider related conflicts that arise even for black holes that do not evaporate significantly.  For the present section, we therefore set aside the original firewall argument of \cite{Almheiri:2012rt} and consider other issues that the reader may consider more or less independent.  The basic tension we explore is one between No Drama and supposing that the black hole is described by a fixed Hilbert space of finite size. Since such a correspondence has been long believed to hold in AdS/CFT, it is useful to frame the discussion in those terms.  This, however is merely a convenience and the fundamental issue transcends this context.

In AdS/CFT there is a sharp dictionary relating the boundary limits of bulk fields to local operators in the CFT.  To extend this further into the bulk requires some form of extrapolation, essentially integrating the bulk field equations.  To extend this past the horizon of a black hole that is formed from collapse, it is necessary to integrate the field equations back in time prior to the formation of the black hole, and then outward to the boundary~\cite{Freivogel:2004rd,Heemskerk:2012mn}.  However, the backwards integration produces an exponential blueshift.  After a time $T^{-1} \ln R$, the backwards integration depends on unknown trans-Planckian interaction  between the Hawking quanta and the infalling body~\cite{Susskind:2012uw}.

This implies that we cannot by this means explicitly construct  the field operators behind the horizon, but a simple argument shows that they do not even exist in principle.  Consider the raising operator $\tilde b^\dagger$ for an interior Hawking mode, which is assumed to have some image in the CFT.  It is an important fact that this {\it lowers} the global energy by some amount $\omega$.  This is possible because the corresponding Killing field becomes spacelike behind the horizon, so the sign of its charge is indefinite.  Now consider all the CFT states with $M < E < M + dM$.   Here $M$ is assumed to be above the Hawking-Page transition, so that typical states behave thermally and have black hole duals, and $dM$ is small but large enough that there are many states in the range.  Labeling these states by
\be
|i\rangle\,, \quad  M < E < M + dM \,,
\ee
consider the states
\be
\tilde b^\dagger |i\rangle \,,  \quad  M - \omega < E < M - \omega + dM \label{raise} \,.
\ee
Now, the raising operator $\tilde b^\dagger$ has a left inverse $\tilde b/(\tilde b^\dagger \tilde b + 1)$ in effective field theory,\footnote{{We are assuming a bosonic field.  For a fermionic field, $\tilde b^\dagger$ would indeed annihilate half of the states, though this still leaves a problem for modes with $e^{-\beta\omega} < \frac{1}{2}$.}}
so the states~(\ref{raise}) must be independent.  However, their number is smaller than that of the $|i\rangle$ by a factor $e^{-\beta\omega}$, so this is a contradiction: the operator $\tilde b^\dagger$ cannot exist in the CFT.

A similar issue clearly arises for any attempt to describe black hole interiors with smooth horizons by a fixed Hilbert space of finite dimension $e^{S_{\rm BH}}$, even in flat spacetime.  For example, the approach discussed in Sec.~3, where nonlocal interactions are added on to the low energy effective field theory of gravity, would have the operator $\tilde b^\dagger$ in its Hilbert space, and so is seemingly inconsistent with a finite $e^{S_{\rm BH}}$.

This result has at least two possible interpretations.  The effective field theory has a UV cutoff, and $\tilde b^\dagger$ may annihilate states at the cutoff.\footnote{This has also been suggested by S. Giddings and J. Maldacena.}  Now, $e^{-\beta\omega}$ is $O(1/2)$, so $\tilde b^\dagger$ annihilates $O(1/2)$ of all states, and $(\tilde b^\dagger)^k$ annihilates {a fraction $1 - O(1/2^k)$} of all states.  With this interpretation, most states are highly excited, near the UV cutoff, and so firewalls are typical.

The other interpretation to explore is that the CFT contains an incomplete description of the black hole interior.  Indeed, the notion that the CFT describes only a subset of the states of the black hole, namely those that could have been formed from collapse, has been expressed before, see Ref.~\cite{Marolf:2008tx} for a discussion.\footnote{{The discussion of \cite{Marolf:2008tx} was based in part on \cite{Maldacena:2001kr,Freivogel:2005qh}.} {Ref. \cite{Gary:2009mi} (and references therein) suggest that even outside horizons the CFT may only describe the bulk coarse-grained over the AdS radius.  We do not agree with this argument.}}  In particular, the situation is similar to the ``bags of gold,'' whose states exceed the Bekenstein-Hawking entropy but which cannot be formed from nonsingular initial data. Since the state created by $\tilde b^\dagger$ is trans-Planckian in the past, there is no guarantee that this state can be formed from collapse, and the counting argument shows that in some cases it cannot.  Of course, an infalling apparatus could emit a quantum in the mode $\tilde b^\dagger$.  However, the mass of the apparatus adds to that of the black hole, and so the full process is more than the creation of the mode.  The  $\tilde b^\dagger$ excitations are also formed by the Hawking process, but always entangled with the $b^\dagger$ {excitations} outside.

On the other hand, an infalling observer who wishes to describe the physics behind the horizon would naturally use low energy effective field theory, including $\tilde b^\dagger$.  Since we for the moment neglect evaporation, we may set aside the concerns of section \ref{AE} and take the point of view of strong complementarity  {(see v1 of \cite{Bousso:2012as}, and \cite{Harlow:2013tf})}, that the external observer can measure $|i\rangle$ but not $\tilde b^\dagger$, and the infalling observer can measure $\tilde b^\dagger$ but not $|i\rangle$.\footnote{Alternately, we might try to define a large Hilbert space containing both, by introducing superselection degrees of freedom \cite{Marolf:2008tx,Marolf:2012xe}.
But this latter is is not sufficient if the corresponding external Hawking modes $b^\dagger$ are fully described by the CFT.  For $b^\dagger, \tilde b^\dagger$ to be part of some effective field theory describing excitations near an infalling vacuum state $|0\rangle$, the state must describe entanglement between states related by $b^\dagger, \tilde b^\dagger$.  In the present context,
$|0\rangle$ would need to describe entanglement between superselection sectors and the CFT.  But as discussed at the end of section \ref{AdSAMPS}, one may form black holes {from collapse} in quantum states with no such entanglement.  So such a scenario would again have firewalls.}

Thus, the nonexistence of $\tilde b^\dagger$ does not in and of itself imply the nonexistence of the interior.  Curiously, if
 $\tilde b$, $\tilde b^\dagger$ did exist in the CFT, we could immediately conclude that typical states would have a firewall.  The point is that the CFT would then contain a full Fock space of states behind the horizon, and so the thermal ensemble would include (and favor, due to their {lower} energy) states of high $\tilde b$ excitation.

However there is yet another counting argument that implies the typicality of the firewall.  Consider the operator $U_\theta = \exp i\int d\omega\,\theta(\omega) N_{b \omega}$.  Here we are using a Schwarzschild frequency basis for the external Hawking modes.  The fields external to the black hole have a mapping to the CFT, and $U_\theta$ commutes with the CFT Hamiltonian to leading order in $1/N$ (free fields in the bulk).  Thus, to good approximation, a high energy state $|i\rangle$ and the the states $U_\theta |i\rangle$ for any $\theta$ are degenerate and so equally weighted in the thermal ensemble.  The operator $U_\theta |i\rangle$ acts  on both the Hilbert space of the asymptotic observer and that of the infalling observer, and so even in the context of strong complementarity we can ask what effect it has on the latter.  The infalling vacuum
is the Bogoliubov transformed state
\be
|0\rangle_A = \exp \left\{ -\int d\omega \, e^{-\beta\omega} \tilde b_\omega^\dagger b^\dagger_\omega \right\}|0\rangle_{\tilde{B},B} \,
\ee
and so
\be
U_\theta |0\rangle_A = \exp \left\{ -\int d\omega \, e^{-\beta\omega + i \theta(\omega)} \tilde b_\omega^\dagger b^\dagger_\omega \right\}|0\rangle_{\tilde{B},B} \,
\ee
is an excited state for nonzero $\theta$.  Equivalently, $U_\theta$ does not commute with the number operator $N_a$ for the
modes of the infalling observer. But $U_\theta$ is invertible and approximately commutes with the Hamiltonian, so the number of states with a given mode excited is at least equal to the number with this mode in vacuum\footnote{{As noted in \cite{Maldacena:2013xja}, the above use of the Schwarzschild frequency basis makes these states singular on the horizon.  This might lead one to worry that our effective field theory treatment is inconsistent.   But one may remove this singularity by replacing the Schwarzschild eigenfrequency modes with similar modes that vanish smoothly at the horizon.  For a large black hole one may keep the commutator $[U_\theta, H]$ small in a regime where effective field theory should remain a valid description of $U_\theta |0\rangle$. }}.  Considering a large number of modes leads to the conclusion that typical states have firewalls.  Even in the context of strong complementarity, this  applies equally to the Hilbert space of an infalling observer  so long as it admits a suitable notion of typical state.

Although this is rather different in form from the original argument of Ref.~\cite{Almheiri:2012rt}, it uses similar assumptions.  That is, we assume effective field theory outside in defining $N_{b\omega}$ and matching it to the CFT.  In place of explicit unitarity of the $S$-matrix, it assumes ordinary quantum mechanics with a finite density of states.

Refs.~\cite{Papadodimas:2012aq,Verlinde:2012cy} have provided recipes to construct fields behind the horizon of an AdS black hole.  The construction is the same as described in \S2, except that $|i\rangle$ is now an arbitrary high energy pure state of AdS, and the CFT is separated into some set of exterior modes $B$ and the rest of the Hilbert space $H$, essentially the black hole interior, which takes the place of $E$. Subsystems will be described by a thermal density matrix, so that
\be
|i\rangle = Z^{-1/2} \sum_{\pmb N} e^{-\beta E_{\pmb N}/2} |i, {\pmb N} \rangle_{H}  |{\pmb N}\rangle_B \,,
\label{entangle}
\ee
for some $|i, {\pmb N} \rangle_{H}$.  By thus using the entanglement with the outside modes, whose AdS/CFT dictionary is known, one can identify arbitrary excited states $|i, {\pmb N} \rangle_H$ of the fields behind the horizon, and also construct operators acting on these.}

In particular, we can construct the operators $\tilde b, \tilde b^\dagger$~(\ref{bbt}) as before.  For fixed $|i \rangle$ there is no conflict with either counting argument above because we are looking only at a small subspace of states.  On the other hand, if we take the union of the construction for all $|i \rangle$, then the counting arguments imply that the states $|i, {\pmb N} \rangle_{H}$ are not independent, and so the action of the $\tilde b, \tilde b^\dagger$ is not well-defined.  Ref.~\cite{Verlinde:2012cy} takes a middle ground of requiring $|i\rangle$ to be restricted to a ``code subspace.''\footnote{Due to the dependence of (\ref{b2}) on $|i\rangle$, one must in fact identify a preferred basis in this subspace.}  This may be consistent with the counting limits, but it is not clear how the code subspace is to be identified.  Is it a fixed subspace determined by the dynamics of the black hole, or is it state-dependent in the fashion described in section 2?

In \S2 we have discussed various problems with this construction, in the context of $\tilde B \subset E$.  {Since in the present non-evaporative context the `fine-grained' Hilbert space is interior to the black hole, some of these issues do not arise. But} the state-dependence of the construction is still problematic.  {Moreover, the proposal is ambiguous}: if we act on an infalling vacuum state with a unitary transformation, we can interpret the resulting state either as an excited state of the original vacuum, or as a vacuum state in its own right.

For example, consider again the operator $U_\theta$.  We have seen that
if  $N_a$ is defined in a state-independent way, in terms of a fixed code subspace, then $N_a e^{i\theta N_b}|i\rangle$ will not vanish, and 
generic values of $\theta$ are not infalling vacuum. On the other hand, if we let $N_a$ be state-dependent, then we are faced with an ambiguity: either view the state
$e^{i\theta N_b}|i\rangle$ as an excited state, or redefine $N_{a}(\theta) = e^{i\theta N_b}N_a e^{-i\theta N_b}$ so that the state is vacuum for any value of $\theta$.  The latter does not agree with the effective field theory for the infalling observer, which contains $N_a$ and $N_b$ as noncommuting operators.

\sect{Varieties of complementarity}

The central question that runs through this subject is the quantum description of the observations of the infalling observer: what  is the nature of his Hilbert space?   We now take an overview of this question.  We begin with the asymptotic observer: at some given time, they can observe the joint state of the black hole $H$, some outgoing Hawking modes $B$ emitted around that time, and the previously emitted radiation $E$.  In an orthonormal basis for $H \otimes B \otimes E$ we have the state $\psi_{i\pmb Nk}$.\footnote{Of course there may also be ingoing matter, but for simplicity we treat $H \otimes B \otimes E$ as a complete description of the physics seen by the asymptotic observer.  To be more precise, however, one should consider $E$ to be the full state of the exterior, aside from the selected modes $B$.}
Now consider an observer who falls into the black hole at around this time.  For their observations we need a density matrix for the inner and outer modes $\tilde B \otimes B$.  {(Actually we need more than this, as we explain at the end of this section, but it is interesting to examine this limited question first).}

The natural way to try to relate these two descriptions is to imagine that $\tilde B$ is identified with some subspace of $H \otimes E$.  Thus we decompose $H \otimes E = \tilde B \otimes \tilde B^c$, and in a basis for $\tilde B \otimes B \otimes \tilde B^c$ the wavefunction is
$
\psi_{\tilde{\pmb N}\pmb N l} .
$
The desired density matrix is then given by the trace over the unobserved degrees of freedom:
\be
\rho_{\tilde{\pmb N} \pmb N,\tilde{\pmb N}{}' \pmb N'} = \sum_l  \psi^*_{\tilde{\pmb N} \pmb Nl}  \psi^{\vphantom{*}}_{\tilde{\pmb N}{}'\pmb N'l} \,.  \label{dm1}
\ee
If we wish to avoid firewalls, this must be pure infalling vacuum.  Thus $\psi_{\tilde{\pmb N} \pmb Nl}$ must factorize as $\phi_{\tilde{\pmb N} \pmb N} \chi_l$, where
\be
\phi_{\tilde{\pmb N} \pmb N} = Z^{-1/2} e^{-\beta E_{\pmb N}{/2}} \delta_{\tilde{\pmb N} \pmb N} \,.  \label{vacdm}
\ee
Now, for {any fixed state of the black hole we can find an identification of $\tilde B \subset H \otimes E$ for which this is true.  However, as we vary over the black hole state the necessary identification changes, being related by some generic unitary transformation.}  Thus the construction~(\ref{dm1}) does not avoid a firewall, unless we extend the rules to allow the embedding of $B \otimes \tilde B^c$ to depend on the state of the black hole.  This is a part of the state-dependence that we have discussed in \S2 and \S5, and which seems to play an essential role in the constructions of Refs.~\cite{Papadodimas:2012aq,Verlinde:2012cy}. Another part is that, even when $\phi_{\tilde{\pmb N} \pmb N}$ is pure, it will not generally agree with the fixed ($\psi_{\tilde{\pmb N} \pmb Nl}$-independent) definition~(\ref{vacdm}) of the infalling vacuum.
The required state-dependence goes beyond the usual rules of quantum mechanics, replacing the density matrix~(\ref{dm1}) with some more nonlinear expression, and so represents a modification of quantum mechanics, whose consistency requires careful consideration.  We will reiterate below a general argument that no such construction can work.

Refs.~\cite{Nomura:2012sw,Nomura3} propose a construction somewhat more elaborate than (\ref{dm1}).  As we have described at the end of \S2, in the model of~\cite{Nomura3} the specific entanglement of $\tilde B$ (there called $C$) with $B$ depends on the specific state in $E$, the so-called classical world.  Thus, the interior mode operator is of the form
\be
\tilde b = \sum_a P^{(a)} \tilde b^{(a)} \,,
\ee
where the $P^{(a)}$ are projectors acting on the early radiation, and the $\tilde b^{(a)}$ are different operators acting on the black hole Hilbert space.  For each classical world there exists a  transformation $U^{(a)}$ acting on $H = \tilde B^{\hat c} \otimes \tilde B$ such that the wavefunction for a classical world $\psi_{h \tilde{\pmb N} {\pmb N} a}$ in $\tilde B^{\hat c} \otimes \tilde B\otimes B \otimes E$ is transformed to the factorized form
\be
\tilde \psi_{h \tilde{\pmb N} {\pmb N} a}^{\vphantom{(a)}} = {\sum_{ h' , \tilde{\pmb M}}} U^{(a)}_{h \tilde{\pmb N},h' \tilde{\pmb M}} \psi_{h' \tilde{\pmb M} {\pmb N} a}^{\vphantom{(a)}}  = \chi_{ha}^{\vphantom{(a)}}  \phi_{ \tilde{\pmb N} {\pmb N} }^{\vphantom{(a)}}  \, .  \label{nomtrans}
\ee
Refs.~\cite{Nomura:2012sw,Nomura3} propose that the state seen by the infalling observer is $\tilde \psi_{h \tilde{\pmb N} {\pmb N} a}$, which gives a pure density matrix for $B\otimes \tilde B$.
Again, this suffers from the same state-dependence as above, (and the more general problem to be described below).  {In particular~(\ref{nomtrans}) is not invertible and thus, in spite of appearances, is not actually a unitary} transformation on the space of states of the black hole:  for different black hole states one needs different $U$'s.


This example is in the framework of an overall Hilbert space, in which the internal Hilbert space is embedded.  Now let us consider strong complementarity.   Here we construct again a density matrix $\rho_{\tilde{\pmb N} \pmb N,\tilde{\pmb N}' \pmb N'}$, which at least for a black hole that forms from collapse should be determined by $\psi_{i\pmb N k}$,\footnote{One might also consider more general forms of complementarity where $\rho_{\tilde{\pmb N} \pmb N,\tilde{\pmb N'} \pmb N'}$ is not fully determined by $\psi_{i\pmb N k}$, but also depends on the history of the black hole.  That is, the black hole would have additional internal degrees of freedom, not represented by the microstate $i$ seen by the exterior observer, which record its history.  However, if we consider black holes formed by collapse, at least in Minkowski spacetime, we can deduce the initial state from the present $\psi_{i\pmb N k}$ from the dynamics.} but where $\tilde B$ is not considered as embedded in $H \otimes E$.  {In this framework one can find a density matrix on $B\tilde B$ with a number of good properties:
\be
\rho_{\tilde{\pmb N} \pmb N,\tilde{\pmb N'} \pmb N'} = \phi_{\pmb N \tilde{\pmb N}}\phi_{\pmb N' \tilde{\pmb N{}'}}
+ \tau_{\tilde {\pmb N}\tilde {\pmb N}{}'} \left( \sum_{ik} \psi^*_{i\pmb N k} \psi^{\vphantom*}_{i\pmb N'k} - \tau_{\pmb N\pmb N'}
\right) \,,  \label{dmstrong}
\ee
where
\be
\tau_{ \pmb N\pmb N'} = Z^{-1} e^{-\beta{E_{\pmb N}}} \delta_{ \pmb N\pmb N'}
\ee
is the thermal density matrix.
First, this is sesquilinear in $\psi$ (with an implicit $\|\psi\|^2$ in two terms), as required by the linearity of quantum mechanics.
Second, summing on $\tilde {\pmb N} = \tilde {\pmb N}{}'$, the reduced density matrix on $B$ for arbitrary $\psi$ is the same for the infalling observer as for the asymptotic observer.  And third, for $\psi$ typical in the microcanonical ensemble, the difference in parentheses vanishes and the infalling density matrix is pure vacuum: there is no firewall.  Unfortunately, this is not positive for general $\psi$.  In particular, if we consider $\psi$ that have been projected along some subspace of $B$, then in the subspace of $B\tilde B$ that is orthogonal to both the projection and $\phi$, only the negative definite $\tau\tau$ term survives.  We believe that one cannot improve on this, but exhibit it as a possibly useful expression.}

Finally, however, there is a general problem with all such constructions.  The density matrix for $\tilde B \otimes B$ does {\it not} describe all infalling observations.  As we have emphasized in point 1 of \S2, the infalling observer can still receive messages from someone who makes measurements on the early radiation.  Thus, a complete description of the infalling observer will involve a density matrix on $\tilde B \otimes B \otimes E$.
Agreement of the $B \otimes E$ density matrix with that of the exterior observer, plus unitarity, plus strong subadditivity, thus imply a firewall.  This is the original argument of Ref.~\cite{Almheiri:2012rt}, and it applies to all the constructions above.

\sect{Conclusions}

We have addressed a variety of possible alternatives to the firewall advocated in \cite{Almheiri:2012rt}.
Guided by AdS/CFT, we have focussed on alternatives that preserve unitarity.  At least in their current forms, these proposals are inadequate.  Proposals to associate the interior of the black hole with the early radiation are highly problematic. Proposals to avoid high energy excitations by introducing non-local interactions at a large distance scale require an implausible conspiracy between such interactions and the dynamics of any mining equipment. We also developed the use of large but evaporating AdS black holes as a context to sharpen the firewall argument.  In this context, fluctuations of the background fields remain small and computational complexity is no obstacle.

Issues associated with black hole evaporation form the sharpest arguments for firewalls.  But \S5 also argued that, even in contexts where the black hole does not evaporate, for generic black hole states the No Drama hypothesis is inconsistent with the black hole interior being described by a fixed Hilbert space of dimension $e^{S_{BH}}$.  We developed two counting arguments, one based on the interior operator $\tilde b^\dagger$ and another based on the exterior operator $U_\theta$.

We again conclude that EFT and Purity require infalling observers to experience high drama at the would-be horizon, at least for sufficiently old black holes.  The detailed form of this drama is beyond the scope of our work.  Thus the `firewall' may not necessarily resemble freely streaming radiation.  For example, it could instead take the form of the large tidal forces that would naively seem to be present in the so-called fuzzball solutions advocated in {\cite{Mathur:2009hf}}
 to replace old black holes.   So such `hard' fuzzballs could perhaps be our firewalls.\footnote{Other proposals that lead to similar behavior include \cite{Chapline:2000en,Mazur:2001fv,Winterberg,Davidson:2011eu,Giveon:2012kp}.  All such suggestions may be categorized as (perhaps suitable limits of) the massive remnant scenario described in \cite{Giddings:1992hh}.}  However, we emphasize that {the arguments above and in \cite{Almheiri:2012rt} limit the extent to which any `fuzzball complementary' proposal along the lines of {\cite{Mathur:2012jk}} can apply to physics inside the horizon.}

The firewall scenario replaces the would-be smooth horizon of a sufficiently old black hole with drama sufficient to remove the entanglement between a Hawking particle and its partner.  Yet the AdS/CFT correspondence implies that radiation emitted by a large AdS black hole remains thermal even for generic states in equilibrium, when the firewall should certainly be active. There is no tension here; to the extent that the firewall remains hidden at or behind the horizon causality prevents it from affecting the radiation outside\footnote{Of course, even in equilibrium the radiation outside must differ in detail from Hawking's prediction over exponentially long times \cite{Maldacena:2001kr}.}.

However, some tension does arise when the black hole evaporates as in section \ref{inAdS}.  While the precise location of the firewall is unclear in a truly dynamical context, this evaporation is sufficiently adiabatic that one expects at each time the firewall to reside at the would-be stationary horizon of the corresponding non-evaporating black hole.  Thus the early part of the firewall could become visible outside as evaporation proceeds.

AdS/CFT indicates that the radiation outside nevertheless agrees broadly with that predicted by Hawking.  {It may be that the firewall should be thought of as the stretched horizon, a dynamical membrane that absorbs and reemits information, but without that additional property that an infalling observer can pass through it.}

This result deserves to be understood in detail, a task far beyond the scope of this work.  However, we see at least two broad scenarios that might explain this result.  First, under the assumption that localizing the firewall to less than a Planck distance may be meaningless, in the above scenario the firewall may perhaps best be considered to reside a Planck distance inside the (adiabatic) horizon rather that precisely at this horizon.  With this interpretation it would naturally fall inward, toward the singularity, faster than the actual event horizon would shrink.  It would thus remain invisible outside, and in fact would need be dynamically regenerated on each fast-scrambling timescale.  Second, it may simply be that the success of the Hawking calculation results from deeper truths about the statistical mechanics of black holes that remain valid even in the presence of firewalls.  {This would be manifestly the case in AdS/CFT under the scenario suggested in \cite{Marolf:2012xe} where generic states of a single CFT have firewalls but, by making use of a different bulk superselection sector, the thermofield double state of two CFTs can also describe a firewall-free two-sided black hole.}

An issue that runs through this whole discussion is the need for a nonperturbative construction of gravity in the bulk.  In our current state of knowledge, we have a nonperturbative construction of the CFT in AdS/CFT, and only a variety of approximate constructions of the bulk.  It is a challenge to do better because of the difficulty of forming precise observables away from the boundary, but this would seem to leave us without any theory with which to understand the dynamics of the firewall.  The considerations in this paper show that if firewalls are to be avoided, it will require some significant departure from our usual framework for physics, likely modifying quantum mechanics for the infalling observer.  One of the supposed lessons of AdS/CFT duality is that quantum mechanics is unmodified, but this applies directly only to bulk observables that have a {\it sharp} CFT dictionary, and these all relate to the AdS boundary.

\section*{Acknowledgements}
We thank Tom Banks, Raphael Bousso, Samuel Braunstein, Steve Giddings, Daniel Harlow, Sabine Hossenfelder, Ted Jacobson, Shamit Kachru, Juan Maldacena, Samir Mathur, Amos Ori, Don Page, Eva Silverstein, Mark Srednicki, Andy Strominger, Lenny Susskind, David Turton, Herman Verlinde, Erik Verlinde, Bob Wald, Aron Wall and the participants of The Nordita Workshop ``The Holographic Way," the Stanford ITP Firewalls meeting, and the CERN TH-Institute ``Black Hole Horizons and Quantum Information,'' for interesting discussions.
AA, JS, and JP were supported in part by NSF grants PHY05-51164 and PHY11-25915, and by FQXi grant RFP3-1017.  DM was supported in part by the National Science Foundation under Grant No PHY12-05500,  by FQXi grant RFP3-1008, and by funds from the University of California.  DS was supported by a KITP Graduate Fellowship, and by the NSF GRF program.


\end{document}